\documentclass[superscriptaddress,floats,showpacs,10pt]{revtex4-1}

\usepackage{graphicx,epsfig}
\usepackage{amsfonts,amsmath,amssymb,amsthm,amscd}
\usepackage[T2A]{fontenc}
\usepackage{color}
\usepackage{dcolumn}
\usepackage{bm}
\usepackage{array}
\usepackage{float}
\begin{document}
\title{Enhancement of maximum attainable ion energy in the radiation pressure acceleration regime using a guiding structure}

\author{S. S. Bulanov}
\affiliation{University of California, Berkeley, California 94720, USA}

\author{E. Esarey}
\affiliation{Lawrence Berkeley National Laboratory, Berkeley, California 94720, USA}

\author{C. B. Schroeder}
\affiliation{Lawrence Berkeley National Laboratory, Berkeley, California 94720, USA}

\author{S. V. Bulanov}
\affiliation{Kansai Photon Science Institute, JAEA, Kizugawa, Kyoto 619-0215, Japan}
\affiliation{Prokhorov Institute of General Physics, Russian Academy of Sciences, Moscow
119991, Russia}
\affiliation{Moscow Institute of Physics and Technology, Dolgoprudny, Moscow region, 141700 Russia}

\author{T. Zh. Esirkepov}
\affiliation{Kansai Photon Science Institute, JAEA, Kizugawa, Kyoto 619-0215, Japan}

\author{M. Kando}
\affiliation{Kansai Photon Science Institute, JAEA, Kizugawa, Kyoto 619-0215, Japan}

\author{F. Pegoraro}
\affiliation{Physics Department, University of Pisa, Pisa 56127, Italy}

\author{W. P. Leemans}
\affiliation{University of California, Berkeley, California 94720, USA}
\affiliation{Lawrence Berkeley National Laboratory, Berkeley, California 94720, USA}

\date{\today}

\begin{abstract}
Radiation Pressure Acceleration relies on high intensity laser pulse interacting with solid target to obtain high maximum energy, quasimonoenergetic ion beams. Either extremely high power laser pulses or tight focusing of laser radiation is required. The latter would lead to the appearance of the maximum attainable ion energy, which is determined by the laser group velocity and is highly influenced by the transverse expansion of the target. Ion acceleration is only possible with target velocities less than the group velocity of the laser. The transverse expansion of the target makes it transparent for radiation, thus reducing the effectiveness of acceleration. Utilization of an external guiding structure for the accelerating laser pulse may provide a way of compensating for the group velocity and transverse expansion effects.                     
\end{abstract}

\pacs{52.25.Os, 52.38.Kd, 52.27.Ny} \keywords{ion accelerators, plasma light propagation, radiation pressure, relativistic plasmas} 
\maketitle

{\noindent} Laser acceleration of charged particles is conceived to be one of the main applications of many powerful laser facilities that are being projected, built, or already in operation around the world. Ultrashort electromagnetic pulses provided by these facilities are able to generate very strong accelerating fields in a plasma, which exceed those of the conventional accelerators by orders of magnitude. This potentially opens a way for compact or even table-top future accelerators providing beams of charged particles ranging from several MeV to multi GeV for many applications. Many research groups focus on laser electron acceleration \cite{Review}, as well as on laser acceleration of ions \cite{ELI,review}. The laser accelerated ion beams can be used in fast ignition \cite{FI}, hadron therapy \cite{hadron therapy}, radiography of dense targets \cite{radiography}, and injection into conventional accelerators \cite{injection}.

There is a wide variety of mechanisms of laser ion acceleration depending on the design of laser matter interaction, ranging from solid density foils to clusters and gas targets, from long to ultra-short pulses, and from $10^{18}$ W/cm$^2$ to $10^{22}$ W/cm$^2$ peak laser intensities. The theoretical studies of the laser ion acceleration show that the Radiation Pressure Acceleration (RPA) \cite{RPA} is one of the most efficient mechanisms of acceleration \cite{RPA, COMREN, OKl}. Several recent experiments may indicate the onset of RPA \cite{RPA exp}. This mechanism is based on the relativistic mirror concept: the laser pulse is reflected back by the co-propagating mirror. The role of the mirror is played either by an ultra-thin solid density foil or by plasma density modulations emerging when the laser pulse interacts with an extended under-critical density target, the so called hole boring RPA \cite{hole-boring RPA, slow wave}. 

In the course of the laser interaction with the mirror there is a momentum transfer from the pulse to the mirror, which results in the downshift of the reflected radiation by a factor of $(1-\beta^2)/(1+\beta^2)\approx 1/4\gamma^2$ for $\gamma\gg 1$, where $\beta$ is the velocity of the mirror, and $\gamma$ is the correxponding to this velocity Lorentz factor. The energy transferred to the mirror can be estimated as $(1-1/4\gamma^2)\mathcal{E}_L$ for $\gamma\gg 1$, where $\mathcal{E}_L$ is the energy of the laser pulse. However, the effects of the electromagnetic (EM) wave group velocity being smaller than the vacuum speed of light, were not taken into account when deriving the scaling for the RPA mechanism. Such effects play an important role in laser-driven electron acceleration \cite{Review} 
and should naturally modify the RPA \cite{slow wave}, especially in the case of tightly focused laser pulses. 
The frequency downshift of such an EM wave reflected by a receding relativistic mirror is \begin{equation}
\omega_{r}=\omega(1-2\gamma^2\beta(\beta_g-\beta)), 
\end{equation}
where $\omega$ and $\omega_r$ are the frequency in the incident and reflected EM wave, 
$\beta_g=v_g/c$ with $v_g$ being the laser pulse group velocity. The energy transferred from the pulse to the mirror is 
\begin{equation}
\Delta\mathcal{E}\simeq 2\gamma^2\beta(\beta_g-\beta)\mathcal{E}_L. 
\end{equation}
If $\beta=\beta_g$, then there is no interaction of the laser light with the target. The group velocity of the pulse limits the value of the attainable velocity of the foil. 

Another limiting factor for RPA by tightly focused pulses comes from the transverse expansion of the target. As the laser pulse diffracts after passing the focus, the target expands accordingly due to the transverse intensity profile of the laser. Due to this expansion, the areal density of the target decreases making it transparent for radiation and effectively terminating the acceleration. It is known that the inclusion of a finite reflectivity of the foil, greatly affects the effectiveness of the RPA mechanism \cite{reflection,unlimited,optimized RPA}.

In what follows we study the RPA of a thin solid density foil by an EM wave with group velocity less than the speed of light in vacuum, $\beta_g<1$. Such waves naturally appear in the case of focused EM radiation or when EM radiation propagates inside some guiding structure or in a medium. In the ultra-relativistic case the energy of ions tends to $\mathcal{F}_L/n_e l$, where $\mathcal{F}_L$ is the laser pulse fluence (incident laser energy per unit area) and $n_e l$ is the areal density of the foil with $n_e$ being the electron foil density and $l$ being the foil thickness \cite{RPA}. The maximum ion energy is determined by the paek laser fluence, $\max[\mathcal{F}_L]$. As shown below, in the case when $\max[\mathcal{F}_L]/n_e l>\gamma_g=(1-\beta_g^2)^{-1/2}$ the group velocity limits the maximum attainable ion energy to $\gamma_g$. We also study the RPA of a thin foil by a diffracting laser pulse and the termination of the acceleration due to increasing transparency of the expanding foil. We show that these two limitations can be mitigated by the utilization of an external guiding structure: the acceleration inside the self-generated channel in the near critical density (NCD) plasma tends to produce ion beams with higher energies.  

In the RPA mechanism of laser ion acceleration the force acting on a foil is expressed in terms of the flux of the EM wave momentum \cite{RPA,COMREN}, which is proportional to the Pointing vector, $\mathbf{S}=\mathbf{E}\times \mathbf{B}/4\pi$. For a circularly polarized wave the vector potential is $\mathbf{A}=A_0(\mathbf{e}_y\cos\varphi +\mathbf{e}_z\sin\varphi )$, $\varphi=\omega t-k x$, where $k$ is the wave vector, the Pointing vector is $\mathbf{S}=\omega k A_0^2 \mathbf{e}_x$. In a frame of reference moving with the foil, the product of wave frequency, $\overline{\omega}$, and wave vector, $\overline{k}$, is given by \cite{slow wave}
\begin{equation}
\overline{\omega}\overline{k}=\omega^2\frac{(\beta_g-\beta)(1-\beta\beta_g)}{1-\beta^2}. 
\end{equation}
In this reference frame the sum of the EM wave fluxes give rise to the force acting on the foil: $(1+|\rho|^2-|\tau|^2)\mathbf{S}$, where $\rho$ and $\tau$ are the reflection and transmission coefficients of the foil in the rest frame of reference. These coefficients enter the energy conservation relation: $|\rho|^2+|\tau|^2+|\alpha|^2=1$, where $\alpha$ is the absorption coefficient. Using these relationships we can write the equation of motion for the on-axis element of the foil, which depends on the peak fluence to obtain the maximum ion energy \cite{slow wave}:
\begin{equation}\label{eq of motion}
\frac{d\beta}{d(\omega t)}=\kappa\beta_g(1-\beta^2)^{1/2}(\beta_g-\beta)(1-\beta\beta_g),
\end{equation}
where 
\begin{eqnarray}
\kappa=(2|\rho|^2+|\alpha|^2)\frac{\omega A_0^2}{4\pi n_e l m_i}\nonumber \\=\frac{1}{2}(2|\rho|^2+|\alpha|^2)\frac{m_e}{m_i}
\frac{a^2(\varphi)}{\varepsilon_e}.
\end{eqnarray}
Here $a=eA/m_e$ is the normalized laser pulse amplitude, $\varepsilon_e=\pi (n_e l/n_{cr}\lambda)$ 
is the parameter governing the transparency of the thin solid density target \cite{Vshivkov}, $n_{cr}=m_e\omega^2/4\pi e^2$ is the critical plasma density, $e$ and $m_e$ are the electron charge and mass respectively, $n_e$ is the electron density in the foil, and $m_i$ is the ion mass. Equation (\ref{eq of motion}) can be solved in quadratures. It reads
\begin{eqnarray}
\left\{\ln\frac{(1-\beta\beta_g+(1-\beta_g^2)^{1/2}(1-\beta^2)^{1/2}\beta_g}{(\beta_g-\beta)(1+(1-\beta_g^2)^{1/2})}\right. \nonumber \\ 
\left.-\beta_g\left[\arctan\frac{(1-\beta_g^2)^{1/2}(1-\beta^2)^{1/2}}{\beta_g-\beta}-\arccos\beta_g
\right]\right\} \nonumber \\=\beta_g(1-\beta_g^2)^{3/2}K_\beta(t) ,  \, \, \, \,
\end{eqnarray} 
where $K_\beta(t)=\int\limits_0^t \kappa dt^\prime$. If we assume that the EM field is constant, then $K_\beta(t)=\kappa t$, 
and for $t\rightarrow\infty$ only the term with $\ln(\beta_g-\beta)$ survives. In this limit we have
\begin{equation}
\beta=\beta_g-\exp\left(-\beta_g(1-\beta_g^2)^{3/2}\kappa t\right). 
\end{equation}
We see that the maximum ion velocity approaches but never exceeds the group velocity of the laser. 
The energy of the ions is limited by $\gamma_g=(1-\beta_g^2)^{-1/2}$. 

Eqn. (\ref{eq of motion}) can b solved numerically for a Gaussian pulse, with duration $\tau=60$ fs, one-lambda focal spot, $r_0=\lambda$, and, thus, the group velocity $\beta_g=1-1/k^2r_0^2\approx 0.975$, which yields $\gamma_g\approx 3.5$. For a circularly polarized laser pulse interacting with a $0.15\lambda$ thick hydrogen foil, the electron density of the foil is equal to $n_e=400n_{cr}$. The results are shown in Fig. 1, where the dependencies of the maximum ion energy on time for three different values of the laser pulse power (100 TW, 200 TW, 1 PW),  for $\beta_g=1$  and $\beta_g<1$ are shown. For a 100 TW laser pulse, i.e. pulse energy equal to 6 J, the ion energy dependence for $\beta_g=1$ and $\beta_g<1$ are almost indistinguishable, which is due to the fact that $\mathcal{F}_L/n_e l<\gamma_g$ ($\mathcal{F}_L/n_e l\approx 1.7$), i.e. the ion velocity is less than the laser group velocity. For a 200 TW laser pulse, i.e. pulse energy equal to 12 J, $\mathcal{F}_L/n_e l\sim \gamma_g$ ($\mathcal{F}_L/n_e l\approx 4.2$) the ion velocity is of the order of the laser group velocity and one can see that there is a slight difference between two corresponding curves in Fig. 1. For a 1 PW laser pulse (with pulse energy of 60 J) the difference between the cases corresponding to $\beta_g=1$ and $\beta_g<1$ is very significant. While the $(\beta_g<1)$-curve is limited by $\gamma_g$, the $\beta_g=1$ curve goes up to $\mathcal{F}_L/n_e l\approx 16.8$. The last case demonstrates the constraint on maximum ion energy due to the laser pulse group velocity being less than the speed of light in vacuum.   

\begin{figure}[tbp]
\epsfxsize7cm\epsffile{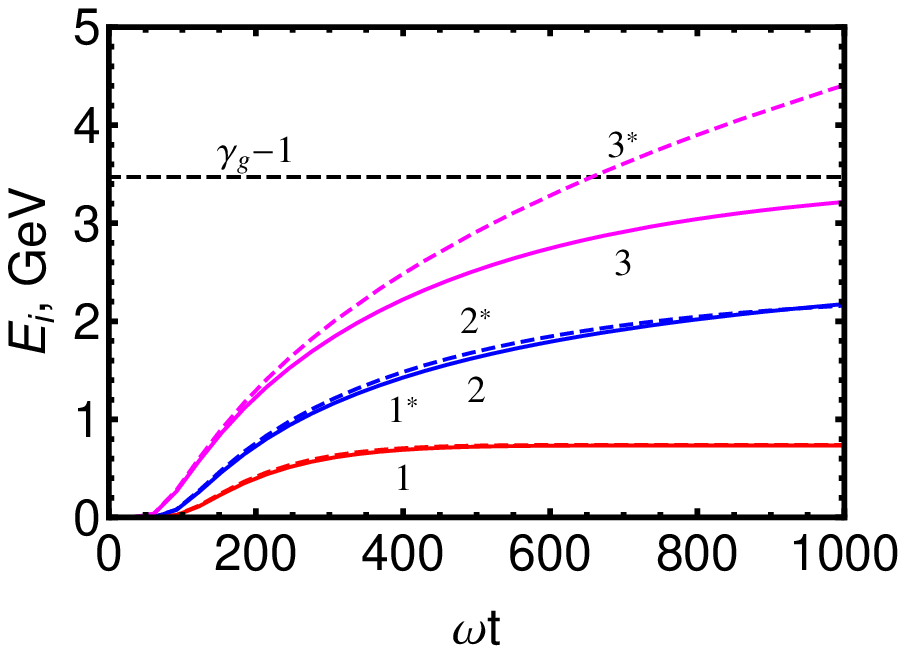}
\caption{The dependencies of the ion kinetic energy on time in case of $\beta_g=1-1/k^2r_0^2$ (solid curves) 
and $\beta_g=1$ (dashed curves) for three different values of the laser pulse power: 100 TW (1), 200 TW (2), and 1 PW (3). 
The density of the foil is $n_e=400 n_{cr}$, the thickness is $l=0.15\lambda$. 
The laser pulse duration is 60 fs, the focal spot is $r_0=\lambda$.}
\end{figure} 

We should mention here that tightly focused pulses diverge rather quickly after passing through the focus. It is plausible to assume that this divergence forces the irradiated part of the foil to expand, following the increase of the laser spot size. If we assume that the field of the pulse can be given by the paraxial approximation, characterized by the laser pulse waist at focus, $w_0$ and the Rayleigh length $L_R=\pi w_0^2/\lambda$, then the evolution of the laser pulse waist as it travels away from focus is $w(x)=w_0\left[1+(x/L_R)^2\right]^{1/2}$, and the amplitude of the field scales with the distance from the focus as $a(x)=a_0\left[1+(x/L_R)^2\right]^{-1/2}$. Since we are interested in the maximum ion energy, we consider RPA of an on-axis element of the foil. The intensity profile near the axis can be approximated by an expanding spherical cup with curvature radius equal to the laser waist, $w(x)$. The on-axis element of the foil can also be approximated by an expanding spherical cup with the curvature $w(x)$ and areal density  equal to $n_e l=n_0l_0\left[1+(x/L_R)^2\right]^{-1}$ and $\varepsilon_e(x) \rightarrow \varepsilon_e(0)\left[1+(x/L_R)^2\right]^{-1}$. Substituting the field and areal density into Eq. (\ref{eq of motion}), we see that the right hand side of Eq. (\ref{eq of motion}) depends on the distance from the focus only through the reflection coefficient. If the foil is opaque for radiation during the entire acceleration process, then the acceleration of an expanding shell by a diverging laser pulse is analogous to the acceleration of an opaque foil, with areal density $n_0l_0$, by an RPA of a plane EM wave with group velocity $\beta_g$ and amplitude $a=a(\varphi)$, 
which depends only on phase $\varphi$.  

In what follows we take into account the decrease of the reflection coefficient as the foil expands 
under the action of the diverging laser pulse and solve Eq. (\ref{eq of motion}) numerically. 
The result of this solution is shown in Fig. 2 by solid curves for laser pulses of 100 TW, 200 TW and 1 PW with $r_0=2\lambda$, the foil thickness of $l=0.15\lambda$ and density of $n_e=400 n_{cr}$. The dashed curves correspond to the solutions with $|\rho|=1$. For 100 TW laser pulse there is almost no difference between the cases with $|\rho|=1$ and $\rho$ being a function of the interaction parameters. This is due to the fact that the motion of the foil is non-relativistic, the displacement of the foil is smaller than the Rayleigh length, and the transverse expansion of the foil is not large enough to affect the reflection coefficient. However, when the motion of the foil becomes relativistic, the displacement of the foil away from the initial position, large enough for the transverse expansion to reduce the reflection coefficient and, consequently, the final energy of the foil. This is illustrated for 200 TW and 1 PW laser pulses in Fig. 2.     

\begin{figure}[tbp]
\epsfxsize7cm\epsffile{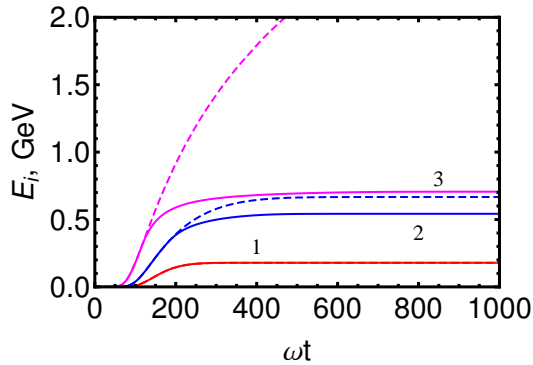}
\caption{The dependencies of the ion kinetic energy on time in case of $\rho=\rho(x,a,\varepsilon_e)$ (solid curves) 
and $\rho=1$ (dashed curves) for three different values of the laser pulse power: 100 TW (1), 200 TW (2), and 1 PW (3). 
The density of the foil is $n_e=400 n_{cr}$, the thickness is $l=0.1\lambda$. 
The laser pulse duration is 60 fs, the focal spot is $r_0=2\lambda$.}
\end{figure} 

\begin{figure}[tbp]
\epsfxsize7cm\epsffile{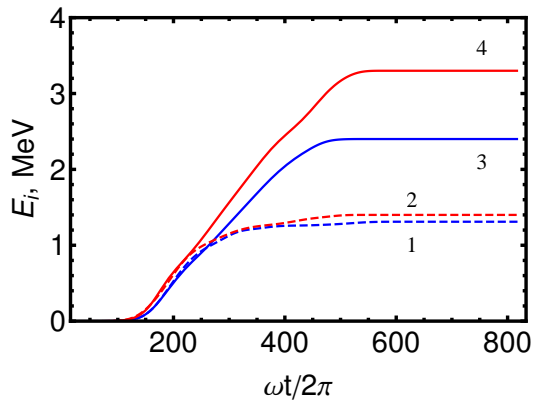}
\caption{The dependence of the ion energy on time for a composite target (solid curves) RPA and single foil (dashed curves) RPA. 
The simulation box is $100\times60\lambda^2$. The laser pulse is initialized at the left border with dimensionless 
potential $a_0$, waist $w$ and duration $\tau$. The pulse is focused at the left front size of the target, 
which is placed 20$\lambda$ away from the left border. The laser parameters at the left border: $a_0=100$, $w=4\lambda$, 
the duration is ten cycles, and $f/D=2$. The composite target consists of a fully ionized hydrogen foil and a hydrogen 
NCD plasma slab placed right behind the foil. The foil thickness is $0.25\lambda$ with densities $n_e=400 n_{cr}$ (curves 1 and 3) and $n_e=225 n_{cr}$ (curves 2 and 4). The thickness of the NCD plasma slab is $50 \lambda$ and density is equal to $n_{cr}$.}
\end{figure}

The utilization of an external guiding structure may relax the limits on maximum attainable ion energy. 
When an intense laser pulse interacts with a composite target, consisting of a thin foil followed by a near critical 
density slab, it accelerates the irradiated part of the foil in the self-generated channel in the NCD plasma. 
Such interaction setup provides less transverse expansion of the target and higher group velocity of the laser pulse. 
In Figs. 3 and 4 we present the results of PIC simulations, obtained by utilizing code REMP \cite{REMP}, which indicate that, for the same laser pulse energy, the ion energy will be significantly larger in the case of a composite target than in the case of a single foil. 

\begin{figure}[tbp]
\epsfxsize4cm\epsffile{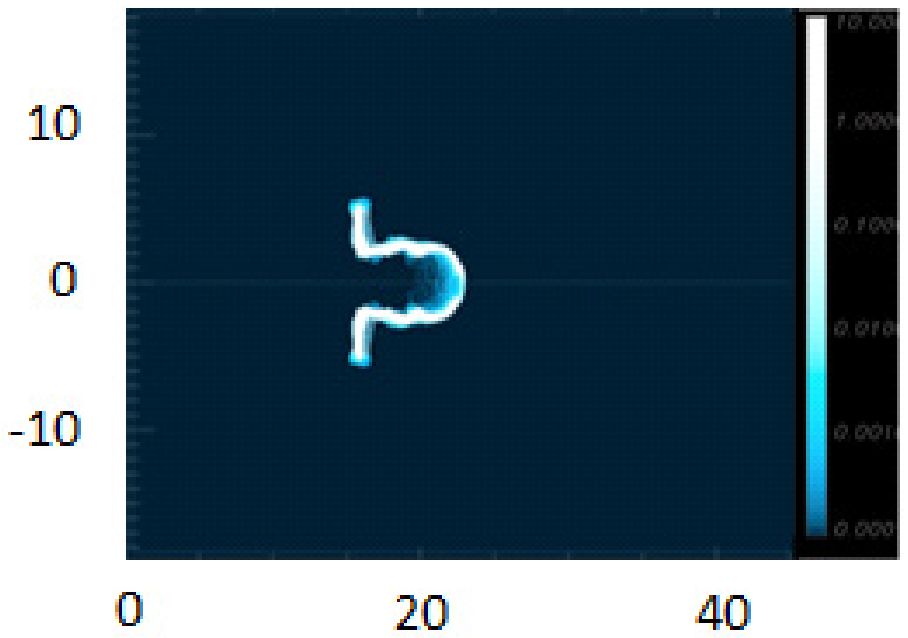}
\epsfxsize4cm\epsffile{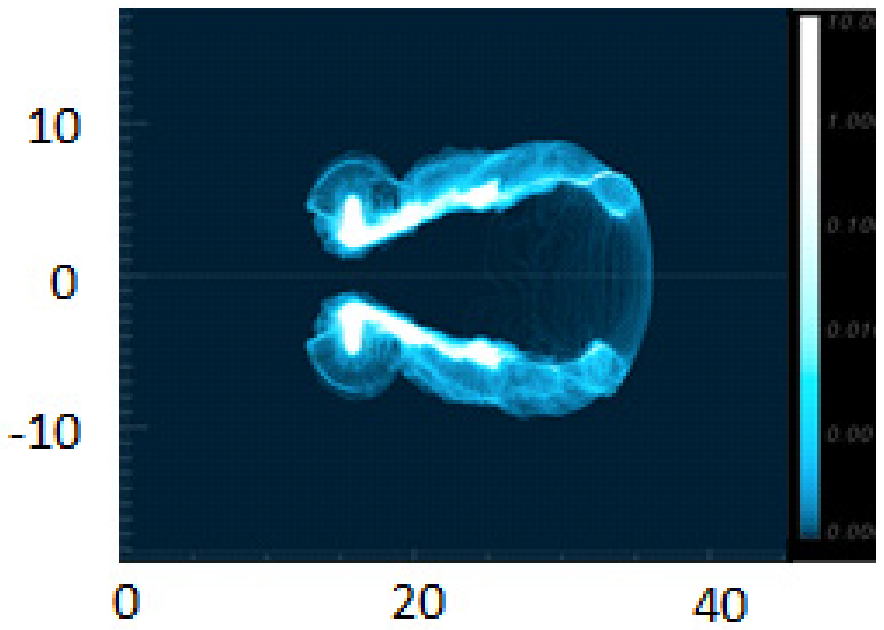}
\epsfxsize4cm\epsffile{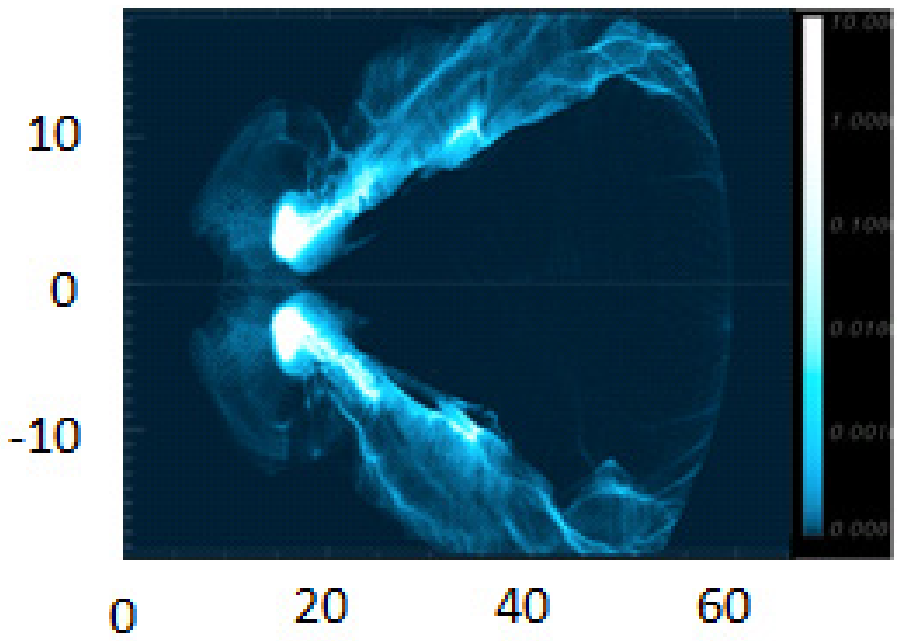}
\epsfxsize4cm\epsffile{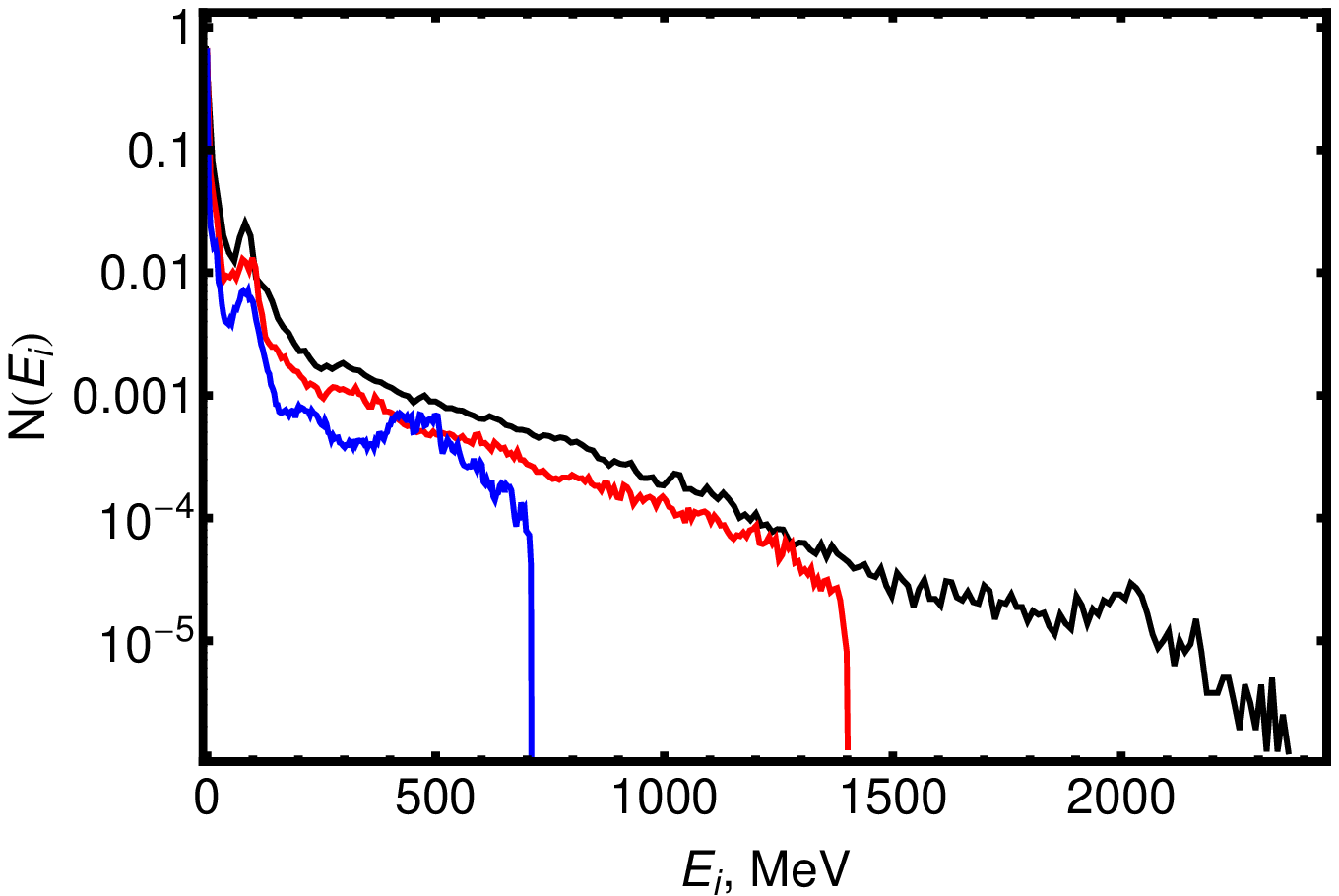}
\caption{The evolution of ion density during the laser pulse interaction with a composite target: a) t=35, b) t=50, c) t=75; 
and the evolution of ion spectrum (c): t=35 (red curve), t=50 (blue curve), and t=75 (black curve). 
The parameters of laser-target interaction are the same as in Fig. 3.}
\end{figure}                      

In conclusion, we showed that a fundamental limitation exists on the maximum attainable ion energy through 
the RPA regime of laser ion acceleration. For a model case of a laser pulse interaction with a non-expanding overdense 
thin foil we showed that the velocity of the target can not exceed the group velocity of the laser pulse. 
Since the RPA mechanism requires relatively high laser intensity to operate, tightly focused laser pulses must be used. 
Such pulses quickly diffract after passing the focus and also force the accelerated target to expand. 
This leads  to the fast decrease of the reflection coefficient which makes the foil transparent to radiation 
and effectively terminates the acceleration. Two main factors have been identified that limit the maximum attainable 
ion energy: the laser pulse group velocity and the transverse expansion of the target due to the acceleration 
by tightly focused laser pulses. The utilization of an external guiding may relax the constraints on maximum attainable ion energy.

We acknowledge support from the NSF under Grant No. PHY-0935197 and the Office of Science of the US DOE under Contract No. DE-AC02-05CH11231 and No. DE-FG02-12ER41798. The authors would like to thank for discussions C. Benedetti, M. Chen, C. G. R. Geddes, and L. Yu.

\end{document}